\documentstyle[eqsecnum,aps]{revtex}

\begin{document}
\draft
\title{Thermalized Displaced Squeezed Thermal States}
\author{Wen-Fa Lu}
\address{CCAST(World Laboratory) P.O. Box 8730, Beijing, 100080,
  \\ and \\
Department of Applied Physics, Shanghai Jiao Tong University,
Shanghai 200030, China
   \thanks{mailing address,E-mail: wenfalu@online.sh.cn}   }
\date{\today}
\maketitle

\begin{abstract}
In the coordinate representation of thermofield dynamics, we investigate the
thermalized displaced squeezed thermal state which involves two temperatures
successively. We give the wavefunction and the matrix element of the density
operator at any time, and accordingly calculate some quantities related to the
position, momentum and particle number operator, special cases of which are
consistent with the results in the literature. The two temperatures have
different correlations with the squeeze and coherence components. Moreover,
different from the properties of the position and momentum, the average value
and variance of the particle number operator as well as the second-order
correlation function are time-independent.
\end{abstract}

Introducing finite temperature effects into squeezed states is important,
because a squeezed state can possesses minimum uncertainty and squeezability
and accordingly technological applicability \cite{1} and, on the other hand, a
finite-temperature influence on it is inevitable. This problem has received
extensive investigations \cite{2,3,4,5,6,7,8,9,10,11}. Analyzing these
investigations, Ref.~\cite{2,3} divided the squeezed states with finite
temperature effects into the thermalized squeezed states \cite{4,5,6} and
the squeezed thermal states \cite{7,8,9,10,11}. The thermalized squeezed states
and the squeezed thermal states are physically distinct states, albeit they are
turned into each other by some parameter transformation \cite{3}. Each of these
two states has its several possible representations, and Ref.~\cite{2} gave a
detailed discussion about them and elucidated their physical interpretations.
From Ref.~\cite{2}, it is not difficult to understand that the squeezed thermal
states correspond to the output from a squeezed device whose input is a thermal
chaotic state with Bose-Einstein distribution, while the thermalized squeezed
states are prepared by thermalizing a squeezed state provided the thermalizing
source is such a device that it can bring a vacuum state into a thermal chaotic
state. Both the thermalized squeezed state and the squeezed thermal state are
involved in one thermal source only. In this paper, we intend to consider yet
another different state which we call the thermalized displaced squeezed
thermal state (TDSTS). This state can be prepared by thermalizing a displaced
squeezed thermal state, and obviously contains the thermalized displaced
squeezed state (TDSS) and the displaced squeezed thermal state (DSTS) as its
special cases. The TDSTS is involved in two thermal sources successively and
perhaps is more practical than the TDSS and the DSTS, because, in general,
there exist thermal noises for the input and output of a displacement-squeeze
device which generates a displaced squeezed state, and furthermore the thermal
noises have perhaps different temperatures for the input and output,
respectively.

In this paper, within the framework of thermofield dynamics \cite{12,13}, we
shall first give the definition of the thermalized displaced squeezed state,
then give the time-dependent wavefunction and calculate the matrix elements of
the density operator in the coordinate representation. From the density matrix
elements, the probability density, average value, variances and the entropy
uncertainty relation of the position and momentum will be calculated. Finally,
we shall also give the average value and variance of the particle number
operator and the second-order correlation function.

This paper is based on the one-dimensional quantum oscillator with a mass $m$
and a constant angular frequency $\omega$ whose Hamiltonian is
\begin{equation}
H={\frac {1}{2m}}p^2 +{\frac {1}{2}} m\omega^2 x^2
  =(a^\dagger a+{\frac {1}{2}})\hbar\omega \;,
\end{equation}
where $x$ is the position operator, $p=-i\hbar{\frac {d}{dx}}\equiv -i\hbar
\partial_x$ the momentum
operator in the coordinate representation, and
\begin{equation}
a={\frac {1}{\sqrt{2m\hbar\omega}}}(ip+m\omega x) \;, \;\;\;\;\;
a^\dagger={\frac {1}{\sqrt{2m\hbar\omega}}}(-ip+m\omega x)
\end{equation}
are the annihilation and creation operators, respectively. Nevertheless, taking
the mass as unit in the formulae of this paper, one can get the results which
are usable for a one-mode electromagnetic field with the same frequency.

As is well known, a displaced squeezed state can be constructed by the squeeze
and displacement operators successively acting on the ground state $|0>$ of the
oscillator Eq.(1) \cite{14}. The displacement operator
\begin{equation}
D(\alpha)=e^{\alpha a^\dagger-\alpha^* a} \;,
\end{equation}
with $\alpha=(\alpha_1+i\alpha_2)=|\alpha| e^{i \gamma}$ any complex number,
corresponds to an ideal displacement device ( the symbol `$|\cdots|$'
represents the module of a complex number ), and the squeeze operator
\begin{equation}
S(z)=\exp\{-{\frac {1}{2}}(z^* a a - z a^\dagger a^\dagger)\} \;,
\end{equation}
with $z=z_1+i z_2=r e^{i\phi}$ any complex number, corresponds to an ideal
squeeze device. ( In Eq.(4), the minus ``$-$'' before ``${\frac {1}{2}}$'' is
sometimes replaced by the plus ``$+$'' in the literature, which gives rise to
no essential differences. ) The action of the displacement operator following
squeeze operator on the ground state will yield the displaced squeezed state
$D(\alpha) S(z)|0>$ ( we call the corresponding device the displacement-squeeze
device ), and a squeezed displaced state $S(z)D(\alpha)|0>$ can be also
constructed by the action of the squeeze operator following the displacement
operator. Note that through a parameter transformation the states
$D(\alpha)S(z)|0>$ and $S(z)D(\alpha)|0>$ can be turned into each other,
because one has \cite{15}
\begin{equation}
S(z)D(\alpha)=D(\alpha \cosh(r)
+\alpha^* e^{i \phi}\sinh(r))S(z) \;.
\end{equation}

In order to consider finite temperature effects, thermofield dynamics \cite{12}
introduces a copy of the physical oscillator Eq.(1) ( called the tilde
oscillator )
\begin{equation}
\tilde{H}={\frac {1}{2m}}\tilde{p}^2 +{\frac {1}{2}} m\omega^2 \tilde{x}^2
  =(\tilde{a}^\dagger \tilde{a}+{\frac {1}{2}})\hbar\omega \;,
\end{equation}
according to the tilde ``conjugation'': $\widetilde{C O}\equiv C^* \tilde{O}$
\cite{12}. Here, $C$ is any coefficient appeared in expressions of quantities
for the physical system, $O$ any operator, the superscript $*$ means complex
conjugation, and $\tilde{O}$ represents the corresponding operator for the
tilde system. Based on the physical and tilde oscillators, thermofield dynamics
manufactures a thermal operator ${\cal T}(\theta)$
\begin{equation}
{\cal T}(\theta)=\exp\{-\theta(\beta)(a\tilde{a}-a^\dagger \tilde{a}^\dagger)\}
                =\exp\{i{\frac {\theta}{\hbar}}(x\tilde{p}- \tilde{x} p)\}
\end{equation}
with $$\tanh[\theta(\beta)]=e^{-\beta\hbar\omega/2}$$
and $\beta={\frac {1}{k_b T}}$. Here, $k_b$ is the Boltzmann constant and $T$
the temperature. The thermal operator is invariant under the tildian
conjugation, $i.e.$, $\tilde{\cal T}(\theta)={\cal T}(\theta)$. Letting
${\cal T}(\theta)$ act on the direct product of the physical ground state $|0>$
and the tilde ground state $|\tilde{0}>$, one can manufacture a thermal vacuum.
Notice that any physical operator commutes with any tilde operator, physical
operators act on physical states only, and similarly tilde operators on tildian
states only. Consequently thermal-vacuum average values in thermofield dynamics
agree with canonical ensemble average values in statistical mechanics \cite{12}.
When the thermal operator acts on the direct product of the free elctromagnetic
field vacuum and its tilde counterpart, one can get a thermal chaotic state
which describes a thermal chaotic light with a Bose-Einstein distribution at
the temperature $T$. That is to say, generally, the thermal operator ${\cal T}
(\theta)$ can correspond to the action of a thermal source.

Now we can give the definition of the TDSTS. Since ${\cal T}(\theta)$ is
involved in the physical and tildian operators simultaneously, we shall use
both the physical operators $D(\alpha)$, $S(z)$ and their tildian counterparts
$\tilde{D}(\alpha)$, $\tilde{S}(z)$ when a finite temperature effect is
introduced into a displaced squeezed state ( The tildian operators
$\tilde{D}(\alpha)$ and $\tilde{S}(z)$ were not adopted in Ref.~\cite{2}).
Because ${\cal T}(\theta)$ commutes with $S(z)\tilde{S}(z)$ \cite{3} and does
not commute with $D(\alpha)\tilde{D}(\alpha)$, we can define a state which
involves two thermal sources with different temperatures successively. Thus, we
have the following TDSTS
\begin{equation}
|\beta_2,\alpha,z,\beta_1,0> \equiv {\cal T}(\theta_2)D(\alpha)
       \tilde{D}(\alpha)S(z)\tilde{S}(z){\cal T}(\theta_1)|0>|\tilde{0}> \;,
\end{equation}
where $\beta_1$, $\theta_1$ and $\beta_2$, $\theta_2$ correspond to those at
the temperatures $T_1$ and $T_2$, respectively. Furthermore, the time evolution
of the TDSTS can be considered by the evolution operator $\hat{U}(t)=
\exp\{-{\frac {i}{\hbar}}(H-\tilde{H})\}$ acting on the TDSTS, $i.e.$,
\begin{equation}
|t,\beta_2,\alpha,z,\beta_1,0>=\hat{U}(t)|\beta_2,\alpha,z,\beta_1,0>
\end{equation}
with $t$ the time. Obviously, the DSTS \cite{2,3,9} is the TDSTS for $T_2=0$,
the TDSS for $T_1=0$, and the thermalized coherent thermal state for $z=0$. The
TDSTS Eq.(8) with proper parameter constraints and transformation can be
reduced into almost all cases discussed in Refs.~\cite{2,3,4,5,6,7,8,9,10,11}
except for the displaced thermalized squeezed states defined by a density
matrix in Ref.~\cite{2}. The TDSTS is physically meaningful. For example, when
a detector examines the output from the displacement-squeeze device whose input
is a thermal chaotic state at the temperature $T_1$, a TDSTS with $T_1$ and
$T_2$ effects will be recognized by the detector provided the detector's added
noise is identified with a thermal chaotic state at the temperature $T_2$. A
thermal chaotic state can be regarded as the ground state with a thermal noise
at some temperature. Thus, the TDSTS involves two thermal noises successively.
For convenience sake, we call the thermal noise at $T_2$ the detector thermal
noise and the thermal noise at $T_1$ the input thermal noise.

In the coordinate representation, the displacement, squeeze and thermal
operators all can be unentangled \cite{16,17}. Moreover, if exploiting the
thermal coordinate representation introduced in Ref.~\cite{17}, one can easily
unentangle the evolution operator $\hat{U}(t)$. Hence one can obtain the
explicit expession of the time-dependent wavefunction
$<\tilde{x},x|t,\beta_2,\alpha,z,\beta_1,0>$. Alternatively, noticing that
$D(\alpha)\tilde{D}(\alpha){\cal T}(\theta)
         ={\cal T}(\theta)D(\alpha(\cosh(\theta)-\sinh(\theta)))
          \tilde{D}(\alpha(\cosh(\theta)-\sinh(\theta)))$
\cite{18}, one has $|t,\beta_2,\alpha,z,\beta_1,0>=
|t,\beta_2,\beta_1,\alpha(\cosh(\theta_1)-\sinh(\theta_1)),z,0>$ and can obtain
easily the time-dependent wavefunction with the help of the result in
Ref.~\cite{17}. Taking $n=0$, $\theta=\theta_1+\theta_2$ and the replacement
$\alpha\to \alpha(\cosh(\theta_1)-\sinh(\theta_1))$ in Eq.(44) of Ref.~\cite{17},
one can read
\begin{eqnarray}
<\tilde{x},x|t,\beta_2,\alpha,z,\beta_1,0>&=&
    \big({\frac {m\omega}{\pi \hbar}}\big)^{\frac {1}{2}}
    {\frac {1}{|{\cal F}_1 B|}}
    \exp\Big\{-{\frac {Q}{B}}-{\frac {Q^*}{B^*}}\Big\}
    \nonumber \\ &\;\;\;& \cdot
    \exp\Big\{-{\frac {m\omega}{2 \hbar}} G_1
    (x \cosh(\Theta)-\tilde{x} \sinh(\Theta))^2
    \nonumber \\ &\;\;\;&
    + 2 \sqrt{{\frac {m\omega}{2\hbar}}} G_2 (\cosh(\theta_1)-\sinh(\theta_1))
 (x \cosh(\Theta)-\tilde{x} \sinh(\Theta)) \Big\}
    \nonumber \\ &\;\;\;& \cdot
    \exp\Big\{-{\frac {m\omega}{2 \hbar}} G^*_1
    (\tilde{x} \cosh(\Theta)-x \sinh(\Theta))^2
    \nonumber \\ &\;\;\;&
   + 2 \sqrt{{\frac {m\omega}{2\hbar}}} G^*_2 (\cosh(\theta_1)-\sinh(\theta_1))
 (\tilde{x} \cosh(\Theta)-x \sinh(\Theta)) \Big\} \;,
\end{eqnarray}
where
\begin{eqnarray*}
 {\cal F}_1&=&\cosh(r)+\sinh(r)\cos(\phi)+i \sinh(r)\sin(\phi), \\
 {\cal F}_2&=&{\frac {1-i \sinh(2r)\sin(\phi)}
    {\cosh(2r)+\sinh(2r)\cos(\phi)}}, \\
\Theta&=&\theta_1+\theta_2, \ \ \ B=\cos(\omega t)+i {\cal F}_2
           \sin(\omega t),\\
  G_1&=&{\frac {{\cal F}_2 \cos(\omega t)+i \sin(\omega t)}{B}},
  G_2={\frac {{\cal F}_2 \alpha_1+i\alpha_2}{B}},
\end{eqnarray*}
and $$ Q=[{\cal F}_2 \cos(\omega t) \alpha_1^2 + 2 {\cal F}_2 \sin(\omega)
 \alpha_1\alpha_2+i \sin(\omega t)\alpha_2^2](\cosh(\theta_1)-
 \sinh(\theta_1))^2 \;.$$
Eq.(10) is the time-dependent wavefunction of the TDSTS in the coordinate
representation, and all information of the TDSTS can be extracted from it.

With the help of Eq.(10), a straightforward calculation yields the density
matrix element $\rho_{x',x}(t)$ of the position for the TDSTS
\begin{eqnarray}
\rho_{x',x}(t)&\equiv& \int^\infty_{-\infty}
           <\tilde{x},x|t,\beta_2,\alpha,z,\beta_1,0>
          <0,\beta_1,z,\alpha,\beta_2,t|x',\tilde{x}> d\tilde{x} \nonumber \\
       &=&\sqrt{{\frac {m\omega}{\pi \hbar}}} {\frac {1}{|{\cal F}_1 B|}}
            \sqrt{{\frac {1}{\cosh(2\Theta)}}}
            \exp\Big\{ {\frac {|{\cal F}_1 B|^2}{2 \cosh(2\Theta)}}
            \coth({\frac {\beta_2\hbar\omega}{4}})(G_2-G^*_2)^2\Big\}  \nonumber \\
  & \ \ \ & \cdot \exp\Big\{-{\frac {m\omega}{4\hbar}}
  {\frac {1}{|{\cal F}_1 B|^2 \cosh(2\Theta)}}
      \Big[x+x'- \sqrt{{\frac {2\hbar}{m\omega}}}
         \sqrt{\coth({\frac {\beta_2\hbar\omega}{4}})}({\frac {\alpha}{A}} +
         {\frac {\alpha^*}{A^*}})\Big]^2
         \nonumber \\   & \ \ \ &
         -{\frac {m\omega}{4\hbar}}{\frac {\cosh(2\Theta)}{|{\cal F}_1 B|^2}}
  \Big[x-x'- \sqrt{{\frac {2\hbar}{m\omega}}}{\frac {|{\cal F}_1 B|^2}
  {\cosh(2\Theta)}}\sqrt{\coth({\frac {\beta_2\hbar\omega}{4}})}(G_2-G^*_2)
  \Big]^2
 \nonumber \\ & \ \ \  &
 -{\frac {m \omega}{4 \hbar}}(G_1-G_1^*)(x^2-x'^2)\Big\}
\end{eqnarray}
with $A=\cos(\omega t)+i \sin(\omega t)$.
This density matrix is complex, Hermitian and time-dependent. When $z=0$
Eq.(11) gives the density matrix of the position for a thermalized coherent
thermal states ${\cal T}(\theta_2)D(\alpha)\tilde{D}(\alpha){\cal T}(\theta)
|0>|\tilde{0}>$ which contains the thermalized coherent state and
the coherent thermal state as its special cases. At the initial time $t=0$ and
$T_2=0$, Eq.(11) is reduced into the density matrix element of the position for
the DSTS
\begin{eqnarray}
\rho_{x',x}&=&\sqrt{{\frac {m\omega}{\pi \hbar}}} {\frac {1}{|{\cal F}_1|}}
            \sqrt{{\frac {1}{\cosh(2\theta_1)}}}
            \exp\Big\{{\frac {|{\cal F}_1|^2}{2 \cosh(2\theta_1)}}
         [({\cal F}_2-{\cal F}^*_2)\alpha_1+i 2\alpha_2]^2\Big\}  \nonumber \\
  & \ \ \ & \cdot \exp\Big\{-{\frac {m\omega}{4\hbar}}
  {\frac {1}{|{\cal F}_1|^2 \cosh(2\theta_1)}}
      \Big[x+x'- \sqrt{{\frac {2\hbar}{m\omega}}}2\alpha_1\Big]^2
      -{\frac {m \omega}{4 \hbar}}({\cal F}_2-{\cal F}_2^*)(x^2-x'^2)
         \nonumber \\   & \ \ \ &
         -{\frac {m\omega}{4\hbar}}{\frac {\cosh(2\theta_1)}{|{\cal F}_1|^2}}
  \Big[x-x'- \sqrt{{\frac {2\hbar}{m\omega}}}{\frac {|{\cal F}_1|^2}
  {\cosh(2\theta_1)}}(({\cal F}_2-{\cal F}^*_2)\alpha_1+i2\alpha_2)\Big]^2
     \Big\} \;,
\end{eqnarray}
which is identical with Eq.(6.5a) in Ref.~\cite{9}(1993). Of course, setting
$T_1=0$, one can obtain the density matrix element of the TDSS. Noticing
\begin{equation}
\cosh(2\Theta)=\coth({\frac {\beta_1\hbar\omega}{2}})
   \coth({\frac {\beta_2\hbar\omega}{2}})
  +{\rm cosech}({\frac {\beta_1\hbar\omega}{2}}){\rm cosech}(
  {\frac {\beta_1\hbar\omega}{2}})
\end{equation}
and $\cosh(\theta)=\coth({\frac {\beta\hbar\omega}{2}})$, one can see that the
differences both among the density matrices of the TDSTS, the DSTS as well as
the TDSS and among the finite temperature influences on these states consist in
the appearance or disappearance of the four factors $\cosh(2\Theta),
\cosh(2\theta_1), \cosh(2\theta_2)$ and
$\coth({\frac {\beta_2\hbar\omega}{4}})$. In the expression of the density
matrices, the factors $\cosh(2\Theta), \cosh(2\theta_1)$ and
$\coth({\frac {\beta_2\hbar\omega}{4}})$ appear for the TDSTS, the factors
$\cosh(2\theta_2)$ and $\coth({\frac {\beta_2\hbar\omega}{4}})$ for the TDSS,
and for the DSTS the factor $\cosh(2\theta_1)$ replaces $\cosh(2\Theta)$.

Taking $x'=x$ in Eq.(11), we have the probability density of the position for
the TDSTS
\begin{eqnarray}
\rho_{x,x}(t)&=&\sqrt{{\frac {m\omega}{\pi \hbar}}}
     {\frac {1}{|{\cal F}_1 B|}}\sqrt{{\frac {1}{\cosh(2\Theta)}}}
     \nonumber \\ & \ \ \ & \cdot
\exp\Big\{-{\frac {m\omega}{\hbar}}{\frac {1}{|{\cal F}_1 B|^2 \cosh(2\Theta)}}
      \Big[x- \sqrt{{\frac {\hbar}{2 m\omega}}}
         \sqrt{\coth({\frac {\beta_2\hbar\omega}{4}})}({\frac {\alpha}{A}} +
         {\frac {\alpha^*}{A^*}})\Big]^2 \Big\}\;.
\end{eqnarray}
This is a Gaussian distribution, and from it one has easily the average value
of the position
\begin{eqnarray}
<x>&\equiv& \int^\infty_{-\infty} x \rho_{x,x}dx
   =\sqrt{{\frac {\hbar}{2 m\omega}}}
         \sqrt{\coth({\frac {\beta_2\hbar\omega}{4}})}({\frac {\alpha}{A}} +
         {\frac {\alpha^*}{A^*}})
         \nonumber \\
   &=&\sqrt{{\frac {2\hbar}{m\omega}}}
         \sqrt{\coth({\frac {\beta_2\hbar\omega}{4}})}
         |\alpha| \cos(\omega t-\gamma)
\end{eqnarray}
and the position variance
\begin{eqnarray}
(\Delta x)^2&\equiv& <x^2>-<x>^2 = {\frac {\hbar |{\cal F}_1 B|^2}{2 m\omega}}
     \cosh(2\Theta)  \nonumber \\
  &=&{\frac {\hbar}{2 m\omega}}[\cosh(2r)+\sinh(2r)\cos(2\omega t-\phi)]
   \cosh(2\Theta)
  \;.
\end{eqnarray}
Here and after, ``$<\cdots>$'' denotes
``$<0,\beta_1,z,\alpha,\beta_2,t|\cdots|t,\beta_2,\alpha,z,\beta_1,0>$''.

Furthermore, exploiting Eq.(11), one can calculate the probability
density of the momentum for the TDSTS
\begin{eqnarray}
\rho_{p,p}(t) &\equiv& <p|\rho|p>=
\int^\infty_{-\infty} {\frac {1}{2\pi \hbar}}
    \exp\{i {\frac {p x'}{\hbar}}-i {\frac {p x}{\hbar}}\} \rho_{x',x} dxdx'
    \nonumber  \\
    & = & \sqrt{{\frac {1}{\pi m\hbar \omega}}}
    \sqrt{{\frac {|{\cal F}_1 B|^2}
    {1-|{\cal F}_1 B|^4 ({\frac {G_1-G_1^*}{2}})^2}}}
 \sqrt{{\frac {1}{\cosh(2\Theta)}}}
    \nonumber  \\
& \ \ \ & \cdot \exp\Big\{-{\frac {|{\cal F}_1 B|^2}
    {m\hbar \omega[1-|{\cal F}_1 B|^4 ({\frac {G_1-G_1^*}{2}})^2]}}
       {\frac {1}{\cosh(2\Theta)}}
    \nonumber  \\
& \ \ \ & \cdot \Big[p-i \sqrt{{\frac {m\hbar\omega}{2}}
          \coth({\frac {\beta_2\hbar\omega}{4}})}
          \Big[{\frac {G_1-G^*_1}{2}}({\frac {\alpha}{A}} +
          {\frac {\alpha^*}{A^*}})-(G_2-G^*_2)\Big]\Big]^2\Big\}
    \nonumber  \\
  &=& \sqrt{{\frac {1}{\pi m\hbar \omega}}}
    \sqrt{{\frac {|{\cal F}_1 B|^2}
    {1-|{\cal F}_1 B|^4 ({\frac {G_1-G_1^*}{2}})^2}}}
 \sqrt{{\frac {1}{\cosh(2\Theta)}}}
    \nonumber  \\
& \ \ \ & \cdot \exp\Big\{-{\frac {|{\cal F}_1 B|^2}
    {m\hbar \omega[1-|{\cal F}_1 B|^4 ({\frac {G_1-G_1^*}{2}})^2]}}
    {\frac {1}{\cosh(2\Theta)}}
    \nonumber  \\
& \ \ \ & \cdot \Big[p+i \sqrt{{\frac {m\hbar\omega}{2}}
          \coth({\frac {\beta_2\hbar\omega}{4}})}
          ({\frac {\alpha}{A}} - {\frac {\alpha^*}{A^*}})\Big]^2\Big\}      \;.
\end{eqnarray}
(Noting that in this paper, some symbols, for example, $p$, $\rho$, should be
understood as either operators or the corresponding eigenvalues of the
operators according to the context where they appear.) The probability density
of the monmentum is also Gaussian. The average value and variance of momentum
are
\begin{equation}
<p> = -i \sqrt{{\frac {m\hbar\omega}{2}}
          \coth({\frac {\beta_2\hbar\omega}{4}})}
          ({\frac {\alpha}{A}} - {\frac {\alpha^*}{A^*}})
    = - \sqrt{2m\hbar\omega
          \coth({\frac {\beta_2\hbar\omega}{4}})}\sin(\omega t-\gamma)
\end{equation}
and
\begin{eqnarray}
(\Delta p)^2 &\equiv& <p^2>-<p>^2=
       {\frac {m \hbar \omega[1-|{\cal F}_1 B|^4 ({\frac {G_1-G_1^*}{2}})^2]}
    {2|{\cal F}_1 B|^2}} \cosh(2\Theta) \nonumber  \\
      &=& {\frac {m\hbar\omega}{2}}[\cosh(2r)-\sinh(2r)\cos(2\omega t-\phi)]
        \cosh(2\Theta)
       \;,
\end{eqnarray}
respectively.

When $T_2=0$ and $t=0$, Eqs.(14) and (17) are consistent with Eqs.(6.6) in
Ref.~\cite{9}(1993). For the case $T_1=0$ and $t=0$ with both $m$ and $\hbar$
unit, if taking the replacement
$\alpha\to [\alpha \cosh(r)+\alpha^*e^{i\phi}\sinh(r)]$ ( Eq.(5) ) in
Eqs.(15),(16),(18) and (19), one can obtain correspondingly the self-tildian
results in Ref.~\cite{4}(1989). Eqs.(15) and (18) indicate that for the TDSTS
the average values of the position and the momentum follows the harmonic motion
of a classical oscillator. The detector thermal noise widens the amplitudes of
the harmonic motion by the factor
$\sqrt{\coth({\frac {\beta_2\hbar\omega}{4}})}$, but the input thermal noise
takes no effects on $<x>$ and $<p>$. However, from the factor $\cosh(2\Theta)$
of Eqs.(16) and (19), the input thermal noise enlarges the variances of the
position and momentum at the same way as the detector thermal noise. It is
worth notice that the joint enlargement factor $\cosh(2\Theta)$ of the
variances by the input and detector thermal noises is greater than the product
of the corresponding enlargement factors for the TDSS and DSTS, namely,
$cosh(2\Theta)> \coth({\frac {\beta_1\hbar\omega}{2}})
\coth({\frac {\beta_2\hbar\omega}{2}})$ ( see Eq.(13) ) .

According to Eqs.(16) and (19), the product of the standard deviations of the
position and momentum is
\begin{equation}
(\Delta x)(\Delta p)={\frac {\hbar}{2}}\cosh(2\Theta)
       \sqrt{\cosh^2(2r)-\sinh^2(2r)\cos^2(2\omega t-\phi)}
\end{equation}
which is consistent with the thermal Heisenberg uncertainty principle
\cite{19}. Obviously, at any time, neither is the TDSTS a minimum Heisenberg
uncertainty state, nor a minimum thermal Heisenberg uncertainty state. This is
different from the TDSS and the DSTS which can be a minimum thermal Heisenberg
uncertainty states when $\cos^2(2\omega t-\phi)=1$.

From Eqs.(14) and (17), one can also obtain the sum of the information
entropies of the position and the momentum in the TDSTS \cite{20}
\begin{eqnarray}
U[x,p|TDSTS]&=&-\int^\infty_{-\infty}\rho_{x,x}(t)\ln[\rho_{x,x}(t)]dx
             -\int^\infty_{-\infty}\rho_{p,p}(t)\ln[\rho_{p,p}(t)]dp
             \nonumber  \\
&=& \ln[\pi e \hbar]+\ln[\cosh(2\Theta)]
  +{\frac {1}{2}}\ln[\cosh^2(2r)-\sinh^2(2r)\cos^2(2\omega t-\phi)]
\end{eqnarray}
which is in conformity with the thermal information-entropy uncertainty
relation \cite{19}. Due to $\cosh(2\Theta)>\cosh(2\theta_2)$, it is impossible
that the TDSTS is a minimum entropy uncertainty state.

The squeeze effect can be discussed by the rotated quadrature phase operators
(the definitions here are slightly different from those in both Ref.~\cite{2}
and Ref.~\cite{14}(the book))
\begin{equation}
Y_1={\frac {1}{2}}(ae^{-i\varphi}+a^\dagger e^{i\varphi}), \ \ \ \ \
Y_2={\frac {1}{2i}}(ae^{-i\varphi}-a^\dagger e^{i\varphi}) \;,
\end{equation}
which give the quadrature phase operators $X_1=
\sqrt{{\frac {m\omega}{2\hbar}}}x$ and $X_2=\sqrt{{\frac {1}{2m\hbar\omega}}}p$
when the rotated angle $\varphi=0$. A straightforward calculation yields
\begin{equation}
(\Delta Y_1)^2\equiv<Y_1^2>-<Y_1>^2={\frac {1}{4}}\cosh(2\Theta)
   [\cosh(2r)+\sinh(2r)\cos(2\omega t+2\varphi-\phi)]
\end{equation}
and
\begin{equation}
(\Delta Y_2)^2\equiv<Y_2^2>-<Y_2>^2={\frac {1}{4}}\cosh(2\Theta)
   [\cosh(2r)-\sinh(2r)\cos(2\omega t+2\varphi-\phi)] \;.
\end{equation}
The last two equations indicate that for the TDSTS, it is always possible to
attenuate either $\Delta Y_1$ or $\Delta Y_2$ at any value of $\phi$ and any
time $t$. At $t=0$, Eqs.(23) and (24) are identical to Eq.(6.9) in Ref.~\cite{9}
(1993) when $\varphi=0$, $T_2=0$, and to Eq.(3.7) in Ref.~\cite{2} when $T_2=0$
and $(\phi-2\varphi)=\pi$.

For the probability densities Eqs.(14) and (17), the moment generating
functions of the position and the momentum are
\begin{equation}
{\cal M}_x(\lambda)\equiv \int^\infty_{-\infty}e^{\lambda x} \rho_{x,x}(t) dx
          =\exp\{\lambda<x>+{\frac {1}{2}}\lambda^2(\Delta x)^2\}
\end{equation}
and
\begin{equation}
{\cal M}_p(\lambda)\equiv \int^\infty_{-\infty}e^{\lambda p} \rho_{p,p}(t) dp
          =\exp\{\lambda<p>+{\frac {1}{2}}\lambda^2(\Delta p)^2\} \;,
\end{equation}
respectively. Hence one can calculate the $n$th moment of the position
as per
\begin{equation}
<x^n>={\frac {d{\cal M}_x(\lambda)}{d\lambda}}\Big|_{\lambda=0}
\end{equation}
 and the $n$th moment of the momentum as per
\begin{equation}
<p^n>={\frac {d{\cal M}_p(\lambda)}{d\lambda}}\Big|_{\lambda=0} \;.
\end{equation}
They can be used to discuss the higher-order squeeze effects for the TDSTS
\cite{21}\cite{9}(1991).

We have discussed the properties related to the position and the momentum. Next,
we will give the average value and variance of the particle number operator
$n=\{{\frac {\hbar}{2 m \omega}}
[({\frac {m\omega}{\hbar}})^2-\partial^2_x] - {\frac {1}{2}}\}$ and the
second-order correlation function. The average value of $n$ at any time $t$ is
\begin{eqnarray}
<n>={\frac {1}{2}}\cosh(2\Theta)\cosh(2r)-{\frac {1}{2}}+
       \coth({\frac {\beta_2\hbar\omega}{4}})|\alpha|^2 \;.
\end{eqnarray}
Thus, the input thermal noise is correlated only with the squeeze, while the
detector thermal noise take effects both on the squeeze and on the coherence.
When $T_2=0$, Eq.(29) is Eq.(3.9) in Ref.~\cite{9}(1993) and Eq.(3.3) in
Ref.~\cite{2}. Exploiting Eqs.(27) and (28), one can calculate the variance of
$n$ at any time and obtain
\begin{eqnarray}
(\Delta n)^2&\equiv& <n^2>-<n>^2
=-{\frac {1}{4}}+
  {\frac {1}{4}}\cosh^2(2\Theta)\cosh(4r)
  \nonumber \\
  & \ \ \ & +\cosh(2\Theta)
 \coth({\frac {\beta_2\hbar\omega}{4}})|\alpha|^2[\cosh(2r)+\sinh(2r)\cos(\phi
  -2\gamma)] \;.
\end{eqnarray}
Consequently one has the second-order correlation function at any time $t$
\begin{eqnarray}
&&g^{(2)}(0)={\frac {<n^2>-<n>}{<n>^2}}
=2+ \nonumber \\ &&
 {\frac {{\frac {1}{4}}\cosh^2(2\Theta)\sinh^2(2r)-
 \coth^2({\frac {\beta_2\hbar\omega}{4}})|\alpha|^4
+\cosh(2\Theta)\coth({\frac {\beta_2\hbar\omega}{4}})|\alpha|^2\sinh(2r)
\cos(\phi-2\gamma)}{[{\frac {1}{2}}\cosh(2\Theta)\cosh(2r)-{\frac {1}{2}}+
       \coth({\frac {\beta_2\hbar\omega}{4}})|\alpha|^2]^2}} \;.
\end{eqnarray}
When $T_2=0$ the last equation is identical to Eq.(3.10) in Ref.~\cite{9}
(1993). From Eqs.(29),(30) and (31), one sees that the average value and
variance of the particle number operator and the second-order correlation
function are time-independent.

In conclusion, this paper has investigated the TDSTS which is a generalization
of the TDSS and the DSTS. In the coordinate representation of thermofield
dynamics, we give the wavefunction and the density matrix with time evolution,
and calculate some quantities related to the position, momentum and particle
number at any time. Our results indicates that the quantities related to the
position and the momentum are time-dependent, but the average value and
variance of the particle number operator and the second-order correlation
function are time-independent. For the influence of the temperature on the
displaced squeezed state, the TDSTS possesses the features both of the TDSS and
of the DSTS, but the finite-temperature effects of the TDSTS are not a simple
sum or product of those of the TDSS and the DSTS. The input thermal noise at
$T_1$ influences only the squeeze via its contribution to the factor
$\cosh(2\Theta)$, while the detector thermal noise at $T_2$ has an additional
effect on the coherence component by the factor
$\coth({\frac {\beta_2\hbar\omega}{4}})$. The TDSTS contains various thermal
squeezed states in the literature as its special cases, except for the
density-matrix-defined displaced thermalized squeezed state in Ref.~\cite{2}.
For the definition Eq.(8), if inserting the thermal operators
${\cal T}(\theta_3)$, ${\cal T}(\theta_4)$ and ${\cal T}(\theta_5)$ between
$D(\alpha)$, $\tilde{D}(\alpha)$, $S(z)$ and $\tilde{S}(z)$, respectively,
without self-tildian requirement (as Ref.~\cite{4}(1989) did), one can obtain a
most generalized displaced squeezed state with finite-temperature effects.
However, we do not know if the most generalized state can be reduced the
density-matrix-defined displaced thermalized squeezed state in Ref.~\cite{2}.
Finally, we want to point out that, if there are $M$ input thermal noises and
$N$ detector thermal noises and the corresponding temperatures are $T_{1,1},
T_{1,2},\cdots, T_{1,M}$ and $T_{2,1}, T_{2,2},\cdots,T_{2,N}$, then the
physical quantities of such a generalized TDSTS can be given by taking
$\Theta=T_{1,1}+T_{1,2}+\cdots+T_{1,M}+T_{2,1}+T_{2,2}+\cdots+ T_{2,N}$ and
$T_2=T_{2,1}+T_{2,2}+\cdots+ T_{2,N}$ in the results of this paper.

\acknowledgments
This project was supported jointly by the President Foundation of
Shanghai Jiao Tong University and the National Natural Science Foundation of
China with grant No. 19875034.

\end{document}